\documentclass[aps,prb,twocolumn,groupedaddress,amssymb,superscriptaddress]{revtex4}

\usepackage[english]{babel}
\usepackage{graphicx}
\usepackage{epsfig}
\usepackage{dcolumn}
\usepackage{amssymb}
\usepackage{amsmath}

\newcommand{\D}{^\dagger}
\newcommand{\PD}{^{\phantom{\dagger}}}

\begin{document}

\bibliographystyle{apsrev}


\title{dc Josephson Effect in Metallic Single-Walled Carbon Nanotubes}

\author{Stefano Pugnetti$^1$, Fabrizio Dolcini$^1$, and Rosario Fazio$^{1,2}$ \\  
${}^1$ {\it Scuola Normale Superiore and   NEST CNR-INFM, I-56126 Pisa, Italy } \\
 ${}^2$ {\it International School for Advanced Studies (SISSA), I-34014 Trieste, Italy}}

\begin{abstract}
The dc Josephson effect is investigated in a single-walled metallic carbon nanotube connected to two superconducting leads. In particular, by using the Luttinger liquid theory, we analyze   the effects of the electron-electron interaction on the supercurrent. We find that in the long junction limit  the strong electronic correlations of the nanotube, together with its peculiar band structure, induce oscillations in the critical current as a function of the junction length and/or the nanotube electron filling. These oscillations  represent a   signature of the Luttinger liquid physics of the nanotube, for they are absent if the interaction is vanishing. We show that this effect can be exploited to reverse the sign of the supercurrent, realizing a tunable $\pi$-junction. 
\end{abstract}

\maketitle

\section{Introduction}

Since the discovery by Iijima in 1991\cite{iijima91},  carbon nanotubes  have attracted much interest in the community of Mesoscopic Physics. Due to their peculiar electronic and mechanical properties, they are regarded as optimal candidates for  nanotechnological implementations, and  have been successfully applied to the realization of quantum transistors\cite{tansetal98,Jarillo-Herrero_Nature_439_953_2006}, electron waveguides\cite{liangetal01}, interferometric devices\cite{liangetal01,Bachtold_nature_397_673_1999} as well as nanoelectromechanical systems\cite{Sapmaz_PRB_67_235414_2003}. 
Recent experiments have spurred the interest in superconducting properties of these materials: it has been observed indeed that proximity-induced superconductivity can arise in nanotube bundles contacted to superconductors~(S); in ropes, intrinsic superconductivity  has also been measured\cite{Kasumov_PRB_68_214521_2003,morpurgoetal99} and  explained in terms of combination of electron coupling to the breathing phonon modes and  intertube Cooper-pair tunneling\cite{Egger_PRB_70_014508_2004}. Individual multiwall nanotubes have recently been utilized in the fabrication of superconductor-nanotube-superconductor hybrid structures, allowing to reveal the dynamics of multiple Andreev reflections\cite{Buitelaar_PRL_91_57005_2003} and to realize a controllable supercurrent transistor\cite{Jarillo-Herrero_Nature_439_953_2006}. By contrast, the investigation of superconducting properties of {\it single-walled} nanotubes in hybrid structures has been only partly explored so far. \\

\begin{figure}
\begin{center}
\includegraphics[angle=90,width=\columnwidth]{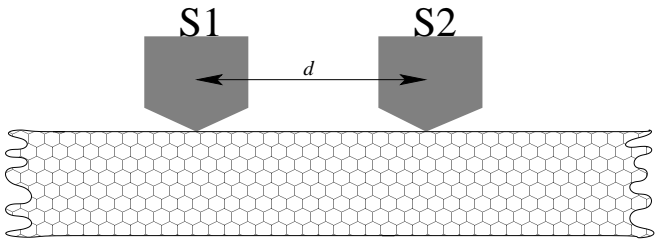}
\caption{\label{fig:1}Schematic set-up of the S-I-SWNT-I-S junction under investigation. }
\end{center}
\end{figure}
Metallic Single-walled carbon nanotubes (SWNT) are known to behave as one-dimensional (1D) conductors with four conduction channels exhibiting ballistic transport up to several $\mu {\rm m}$ \cite{tansetal97,white_nature_393_240_1998}. Differently from other 1D metals, SWNT preserve their conduction properties even at very low temperature, since the cylindrical lattice geometry prevents the arising of Peierls distorsion. They thus offer promising features for interconnecting components of nanodevices.
Due to their 1D character, electronic correlations have dramatic effects on the behavior of SWNT: experimental evidences of a power-law behavior for the conductance as a function of temperature\cite{bockrathetal99} indicate that SWNT exhibit a Luttinger liquid (LL) behavior, and that their elementary excitations are not fermionic quasi-particles like in normal 3D metals\cite{voit95,kanefisher92}. It is thus   expected  that, when a SWNT is contacted to S leads at equilibrium, electronic correlations might significantly modify the behavior of the supercurrent with respect to  junctions realized with a normal metal. This issue has been addressed in the literature \cite{afflecketal00,maslovetal96,cauxetal02,takane96,fazioetal96,fazioetal95} and it has been shown that the effect of interaction is particularly enhanced when the coupling between the LL and the S leads is realized through tunnel junctions. However, most of these investigations focused on the case of a two-channel (i.e.~one spinful mode) LL, and cannot be straightforwardly applied to the case of a four-channel SWNT. 
In this paper we discuss this problem investigating the dc Josephson effect in a S-I-SWNT-I-S junction, and show that new features arise due to the peculiar  band structure of SWNT. The paper is organized as follows: in Sec.~II we briefly review the model used to describe SWNT, accounting for electron-electron interaction within the Luttinger Liquid theory. In Sec.~III we present our results about the Josephson current. We find that the interaction yields a twofold effect on the critical current $j_c$: on the one hand it modifies the scaling law of $j_c$ as a function of the junction length~$d$; on the other hand, it introduces oscillations of $j_c$ as a function of either the electron filling or the junction length~$d$. The latter oscillations are absent for a non-interacting system, and therefore represent a signature of Luttinger liquid behavior on the  supercurrent. Finally, in Sec.~IV we discuss the results and propose possible implementations to observe this effect.\\

\section{Modeling the system}
The set-up of the system is depicted in  Fig.~\ref{fig:1}: a  metallic SWNT is coupled through tunnel contacts to two superconducting leads to realize a S-I-SWNT-I-S junction. For simplicity, we limit here our treatment to the case of armchair nanotubes; we also assume that the S leads have the same gap $|\Delta|$; the two superconducting order parameters thus read $\Delta_{1,2}=|\Delta| e^{i\chi_{1,2}}$, where $\chi_{i}$ is the superconducting phase of the $i$-th lead. We are interested in the dependence of the critical current on the junction length~$d$; we thus analyze the regime
\begin{equation}
\lambda_c \, , \xi \, \ll d \, \ll L \label{regime}
\end{equation}
where $\lambda_c$ represents the width of the contacts, $\xi$~the coherence length of the S electrodes, $d$ the electrode distance, and $L$ the length of the nanotube. The regime~(\ref{regime}) is quite realistic in view of  customary fabrication of  $\mu {\rm m}$ long ballistic nanotubes\cite{white_nature_393_240_1998}, and the recent realization of superconducting tips for Scanning Tunneling Microscope (STM) \cite{panetal98,rodrigoetal04,kohenaetal05} or of 10-20 nm short  superconducting finger leads. In order to simplify the mathematical treatment without altering the  essential physical features of the regime (\ref{regime}), we shall henceforth assume that  the tunnel contacts are point-like, the coherence length $\xi$ is vanishing, and the length of the nanotube is infinite, $L \rightarrow \infty$. \\

\begin{figure}
\begin{center}
\includegraphics[width=\columnwidth]{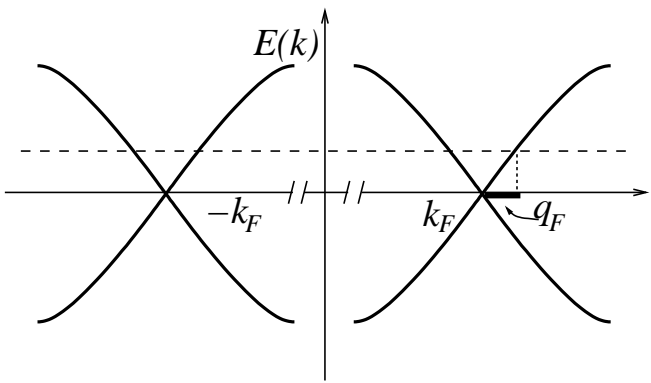}
\caption{\label{fig:2} The electron band dispersion relation of a SWNT originates from the two-sublattices honeycomb carbon structure, and is characterized by four Fermi points. The latter can be identified through two Fermi momenta: $k_F$ denotes the band crossing points, whereas $q_F$ accounts for the deviation from $k_F$, i.e. the electron filling of the SWNT.}
\end{center}
\end{figure}

In a metallic nanotube the lowest band  consists of four electron branches located around two Fermi points $\alpha k_F$, with $\alpha = \pm 1$; the energy separation to the second band is of the order of~${\rm eV }$, so that the latter can be in practice neglected up to a broad range of thermal excitations. Within this energy scale, the energy dispersion of the lower band is linear under quite good approximation, as shown in Fig.~\ref{fig:2}. SWNT can thus be regarded as four-channel 1D metals. As discussed in the introduction, their 1D character implies that a careful treatment of the electron-electron interaction is needed. It is indeed well known that   transport properties of SWNT cannot be explained in terms of the customary Fermi liquid theory, since their elementary excitations   are bosonic plasmon modes, rather than fermionic quasi-particles. A model for SWNT based on the Luttinger liquid theory has been formulated a decade ago\cite{kaneetal97,eggergogolin97}, and applied in a number of problems\cite{kimetal06,lebedevetal05,gaoetal04,pecaetal03,crepieuxetal03,komnikgogolin02,komnikegger01}. Here we briefly remind the main aspects that are relevant to our discussion: A SWNT can be ideally obtained by wrapping into a cylinder shape a graphene sheet, whose honeycomb carbon lattice consists of two sublattices $p = \pm$. A nearest-neighbor tight-binding calculation of the $\pi$-electrons in the graphite, together with appropriate  wrapping boundary conditions, leads to express the electron field in the nanotube  as
\begin{equation}
\label{eq:fullpsi}
\Psi_\sigma(x)= \hspace{-0.3cm} \sum_{\alpha=\pm , r=R/L}  \, \,  \sum_{p=\pm}  {U}_{p r}   e^{i(\alpha k_F+rq_F)x} {\psi}_{\alpha r \sigma}(x) 
\end{equation}
where $\sigma=\uparrow, \downarrow$ denotes the spin component and  $x$  the longitudinal coordinate in the nanotube. In Eq.~(\ref{eq:fullpsi}), $U_{p r}$ are the entries of the matrix
\begin{equation}
U=- \frac{e^{-i \pi/4}}{\sqrt{2}} \left(
\begin{array}{lcr}
1 & 1 \\
i & -i
\end{array} \right)
\end{equation}
describing the unitary transformation from the sublattice electron fields into the right(left) moving  fields description. The exponential terms in Eq.~(\ref{eq:fullpsi}) represent the fast oscillating contribution to the electron wave function, where the wave vector $q_F$ is related to the electron filling exceeding the Fermi points $\pm k_F$, as illustrated in Fig.~\ref{fig:2}. Finally the field ${\psi}_{\alpha r \sigma}(x)$ varies slowly over the scale of Fermi wavelength. 

In order to account for the electron-electron interaction, it is useful to represent the electron fields $\psi_{\alpha r \sigma}$ through the bosonization identity
\begin{equation}
\psi_{\alpha r \sigma}(x)=
\frac{\eta_{\alpha r \sigma}}{\sqrt{2\pi a}}
\exp\{
i\varphi_{\alpha r \sigma}(x)\} \label{bosonization}
\end{equation}
where $\varphi_{\alpha r \sigma}(x)$ is the plasmonic field  describing the long wavelength fluctuations. The operators $\eta_{\alpha r \sigma}$ are Klein factors obeying a Clifford algebra and ensuring the correct anticommutation  between  different fermionic species. Finally $a$ is a cut-off length  regularizing the theory, and is of the order of the lattice spacing.  
The effective hamiltonian for the SWNT reads
\begin{equation}
\label{H-SWNT}
{\mathcal H}_{\rm SWNT}= 
\sum_{j \delta}\frac{v_F} 2
\int_{-\infty}^{+\infty} dx
\left[
 (\partial_x\phi_{j\delta})^2+
  \frac{(\partial_x\theta_{j\delta})^2}{K^2_{j\delta}}
\right]  
\end{equation}
with $v_F\simeq8\cdot10^5$ ms$^{-1}$. Here $\theta_{j \delta}(x)$ are four independent bosonic fields, with  $j$ labeling charge(c) and spin(s) degrees of freedom, and $\delta=\pm$ denoting two independent linear combinations of the $\alpha= \pm$ branches. The fields $\phi_{j \delta}(x)$ are related to $\theta_{j \delta}(x)$ by the duality relation $[ \theta_{j \delta}(x), \partial_y \phi_{j \delta}(y) ] = i \delta(x-y)$, 
and $K_{j \delta}$ are interaction parameters, with $K_{j \delta} < 1$ ($K_{j \delta} > 1$) for repulsive (attractive) interaction and  $K_{j \delta} = 1$ for vanishing interaction. It can be shown \cite{eggergogolin98} that the mode $(j=c, \delta=+)$ is strongly interacting ($K \doteq K_{c +} \simeq 0.3$), while the  three other modes are neutral $K_{j \delta \neq c +} = 1$. The  fields $\varphi_{\alpha r \sigma}$ appearing in the bosonization identity (\ref{bosonization}) are linear combinations of the decoupled modes 
\begin{equation}
\begin{split}
\varphi_{\alpha r \sigma}
=&
\frac{\sqrt{\pi}}2
\left\{
  \phi_{c+}+r\theta_{c+}+\alpha\phi_{c-}+r\alpha\theta_{c-}
\right.\\
&\left.
  +\sigma\phi_{s+}+r\sigma\theta_{s+}+\alpha\sigma\phi_{s-}+r\alpha\sigma\theta_{s-}
\right\}
\end{split}
\end{equation}

The Hamiltonian modeling the S-I-SWNT-I-S junction thus reads
\begin{equation}
{\mathcal{H}}={\mathcal{H}}_{\rm SWNT} \, + \, {\mathcal{H}}_{\rm SC_1} \, + \, {\mathcal{H}}_{\rm SC_2} \, + \, {\mathcal{H}}_{\rm T}
\end{equation} 
where ${\mathcal{H}}_{\rm SWNT}$ is given by (\ref{H-SWNT}), ${\mathcal{H}}_{\rm SC_{1,2}}$ are the usual BCS hamiltonians for the electrodes, and ${\mathcal{H}}_{\rm T}$ describes the nanotube-electrodes  electron tunneling. Denoting by $x_i$ the nanotube coordinate of the injection point to the $i$-th electrode, one can write
\begin{equation}
{\mathcal{H}}_{\rm T} = \sum_{i=1,2} \sum_{\sigma=\uparrow, \downarrow} T_i \left( \Xi^\dagger_i (x_i) \Psi^{}_\sigma(x_i) + \Psi^{\dagger}_\sigma(x_i) \Xi^{}_i (x_i) \right) 
\end{equation}
where $\Xi^\dagger_i $ is the electron field operator in the $i$-th lead, and $T_{1,2}$ are tunneling amplitudes.   The Josephson current is computed perturbatively in the tunneling amplitudes.

\section{dc Josephson Current}
Denoting by $F$ the free energy of the S-I-SWNT-I-S junction, and by $\chi=\chi_2-\chi_1$ the phase difference between the two superconductors, the dc Josephson current is obtained  as
\begin{equation}
I_J=\frac{2e}\hbar\frac{\partial F}{\partial\chi} \, \, .
\end{equation}
Evaluating $I_J$  to the fourth order in the tunneling amplitudes $T_i$, one obtains (up to $\chi$-independent terms)
\begin{eqnarray}
\lefteqn{F =- 2 \frac{(T_1 T_2)^2}{\beta} \Re \left[ \prod_{i=1}^{4} \int_0^\beta\!\!\!d \tau_i   \right.}  \hspace{1cm} \label{free-en}
& \\
& \times \, \left.  \mathcal{F}_1(\tau_1-\tau_2)
\mathcal{G}(\tau_1 ,\tau_2, \tau_3,\tau_4;d) \mathcal{F}^*_2(\tau_3-\tau_4)  \right] \, \, . \nonumber
\end{eqnarray}
In Eq.~(\ref{free-en}) $\beta$ denotes the inverse temperature and
\begin{equation}
\label{eq:anomalouspropagator}
\begin{split}
\mathcal{F}_{i}(\tau - \tau')&=
\langle
  {\rm T} \left\{
  \Xi_{i,\uparrow}\D(x,\tau)\Xi_{i,\downarrow}\D(x,\tau')
  \right\}
\rangle=\\
&=
\frac{\pi N(0)}{\beta}e^{-i\chi_i}\sum_{n\in\mathbb{Z}}e^{-i\omega_n(\tau-\tau')}\frac{|\Delta|}{ \sqrt{\omega_n^2+|\Delta|^2}}
\end{split}
\end{equation}
the anomalous BCS   T-ordered correlator  in the $i$-th S lead, with a density of states of the normal state $N(0)$ at the Fermi energy, and  Matsubara frequencies $\omega_n=(2n+1)\pi/\beta$. Finally
\begin{equation}
\label{Gdef}
\begin{split}
\mathcal{G}(\tau_1&,\tau_2, \tau_3,\tau_4;d)=\\
=&\langle
  {\rm T} \left\{
  \Psi_{\uparrow}\PD(0,\tau_1)\Psi_{\downarrow}\PD(0,\tau_2)
  \Psi_{\downarrow}\D(d,\tau_3) \Psi_{\uparrow}\D(d,\tau_4)
  \right\}
\rangle
\end{split}
\end{equation}
is the two-electron T-ordered correlator in the SWNT.  Under the condition (\ref{regime}), one has $\Delta\gg \hbar v_F/ d$, implying that the lead-nanotube tunneling time is much shorter than the traversal time  $v_F/d$ along the junction. Eq.~(\ref{eq:anomalouspropagator}) is then well approximated by  a $\delta(\tau-\tau')$, and tunneling effectively involves electron pairs. In this regime   the Josephson current can be written as
\begin{equation}
I_J=I_0 (\chi) \, j_c(d;T) \label{I_J}
\end{equation}
\\Here the first term
$ I_0 = ({2e}/{\hbar}) (\hbar v_F/d) \mathcal{T}  \sin \chi$
accounts for the dependence on the superconducting phase difference $\chi$, and corresponds to the  Josephson current of a long ballistic junction with bare transmission coefficient 
$
\mathcal{T}=
(4/2\pi) |T_1T_2|^2\pi^2(N(0)/\hbar v_F)^2
$ at zero temperature. The second term represents the (dimensionless) critical current and
encodes the details of the junction: it  depends  on the length, on  the temperature, and on the interaction   effects, as we shall see below. Explicitly, it reads
\begin{eqnarray}
  j_c(d;T)=
  \frac1{2\pi}
  \left(
    \frac{d}{a} 
  \right)^2
  \int_{-1/\Theta}^{1/\Theta}\!\!d\xi
  \left(
    1-\Theta|\xi|
  \right)\times\hspace{2cm}
\label{jd}\\
\sum_{r=\pm,\alpha\alpha'} 
\!\!\!
\frac{e^{-i(\alpha+\alpha')k_F d}}2
[ C^A_{r \alpha \alpha'} 
  \left(
    k_F d,  \xi
  \right)+
  e^{2ir q_F d}  C^P_{r \alpha \alpha'}
  \left(
    k_Fd,\xi
  \right)
]
\nonumber
\end{eqnarray}
with $\Theta= k_B T d/\hbar v_F$.  Two types of processes, denoted by $P$ and $A$, contribute to $j_c$: the former (latter) describes tunneling of Cooper pairs formed by electrons with parallel (antiparallel) momenta. The related pair operators  
\begin{equation}
\begin{split}
O^{A}_{r,\alpha\alpha'}(x,\tau)=&
\psi_{\alpha r\uparrow}\PD(x,\tau)\psi\PD_{\alpha'-r\downarrow}(x,\tau)\\
O^{P}_{r,\alpha\alpha'}(x,\tau)=&
\psi\PD_{\alpha r\uparrow}(x,\tau)\psi\PD_{\alpha'r\downarrow}(x,\tau)
\end{split}
\end{equation}
yield the two   correlators 
\begin{equation}
\begin{split}
  C_{r \alpha \alpha'}^{P/A}&( k_F d,\xi_3-\xi_1) =\\
  =&
  (2\pi a)^2
  \langle
    T\left\{
      O_{r,\alpha\alpha'}^{P/A}(0,\frac {d}{\hbar v_F}\xi_1) O_{r,\alpha\alpha'}^{\dagger \, P/A}(d,\frac {d}{\hbar v_F}\xi_3)
    \right\}
  \rangle
\end{split}
\end{equation}
appearing in Eq.~(\ref{jd}). Importantly, these two terms correspond to  different  dependences on the  momenta   defining the four Fermi points of the SWNT: while  $A$ processes  only involve $k_F$, $P$ processes are also characterized by the electron filling momentum $q_F$, as can be seen from the phase factors multiplying $C_{r \alpha \alpha'}^{P/A}$. Since typically $q_F \ll k_F$, two  periods are expected to arise in the dependence of the Josephson current on the junction length~$d$.  However, this is not necessarily the case.
In the first instance, indeed, the dependence on $k_F$  amounts to a prefactor $1+\cos(2 k_F d)$, and is extremely difficult to be observed in a realistic system where the approximation of point-like contacts is not valid, for the period of these oscillations is usually smaller than the  typical contact width $\lambda_c$. Even in the regime (\ref{regime}), the observed current is in fact an average $\langle \ldots \rangle_{\lambda_c}$ over lengths $d+x$, where $x$ ranges over $\lambda_c$. This averaging effectively yields  
\begin{equation}
\langle 1+\cos{(2 k_F d)} \rangle_{\lambda_c} = 1
\end{equation}
so that the dependence on $k_F$ disappears. The results for the current presented henceforth  are thus meant upon performing this averaging procedure.  \\
Secondly, the electron-electron interaction strongly affects the behavior of the correlators $C_{r \alpha \alpha'}^{P/A}$. Although the full  expression  for the latter is quite lengthy (see the appendix for details), important insights can   be gained from the analysis of the scaling dimensions of the two operators;  one obtains that for $ k_F d \gg 1$
\begin{equation}
\begin{split}
|C_{r \alpha \alpha'}^{P/A}(k_F d,\xi)|
\sim &
\left|
  \frac{a}{d}
\right|^{2\cdot \delta_{P/A}}
\end{split}
\end{equation}
with $\delta_P=(K+1/K+2)/4$ and $\delta_A=(1/K+3)/4$. While for vanishing interaction ($K=1$) the scaling dimensions of the two processes coincide, the electron interaction modifies the power laws of these two processes in a different way:  the  contribution of $A$ processes decays faster than the one of $P$ processes ($\delta_P < \delta_A$). Remarkably, this does {\it not} imply that for a sufficiently long junction the Josephson current is mainly due to $P$ processes. Indeed, an electron pair traveling along the junction also acquires a phase, which results into oscillating factors in the correlators.  Since the Josephson current (\ref{jd}) depends on the integral over the imaginary time variable $\xi$, not only the decay rate but also the  phase of  $C_{r \alpha \alpha'}^{P/A}$ matters. Since the dynamics of the electrons is coupled by the interaction, these phase factor are also affected by the value of $K$. \\

In the case that electron interaction is neglected ($K=1$), the effect of alternating phases is so strong that the total contribution of $P$ processes vanishes. Indeed, when integrating over all possible pair momenta, the phase acquired by the electron pair traveling along the junction  oscillates,  yielding a cancellation of the different contributions, except for those processes in which the total pair momentum is vanishing. While this condition can be fulfilled by $A$ processes, simple phase-space considerations show that the total contribution of $P$ processes  is suppressed.  As a consequence,   the Josephson current through a SWNT is predicted to be independent of the filling momentum $q_F$, and one obtains
\begin{equation}
j_c=\frac{2 \pi \Theta}{\sinh(2\pi \Theta)}
\end{equation}
At zero temperature, $j_c = 1$, and the Josephson current scales as $1/d$ due to the $I_0$ term (see Eq.~(\ref{I_J})), whereas at finite temperature it is exponentially suppressed.\\

\begin{figure} 
\includegraphics[width=\columnwidth]{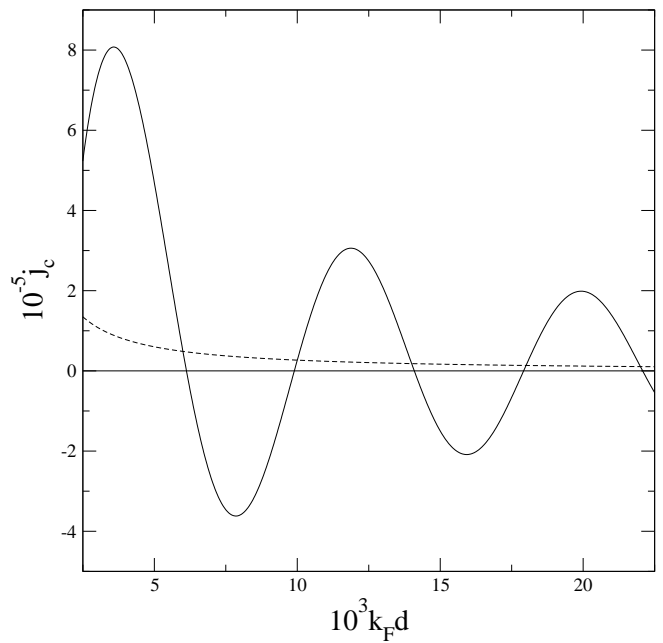}
\caption{\label{fig:3}(solid curve): The oscillations of the dimensionless Josephson current   as a function of junction length $d$ for a S-I-SWNT-I-S junction,  at zero temperature. The SWNT filling factor is $q_F/k_F=10^{-3} \pi/8$ and its interaction parameter is $K=0.3$. The oscillations have  a period $2 q_F d$ and decay with a power-law with an interaction-dependent exponent $\delta_P$. (dashed curve): The contribution of $A$ processes to the critical current is monotonous and positive, indicating that the oscillations originate from $P$ processes (see text).}
\end{figure}
In contrast, when electron-electron interaction is taken into account ($K \simeq 0.3$), the dynamics of the two electrons forming any  pair  is coupled, and the mechanism leading to the cancellation occurring in the non-interacting case is not valid. Electronic correlations both affect the contribution of $A$ processes  and make an oscillating contribution in $2 q_F d$ arise from $P$ processes.   These oscillations are characterized by a much longer period than the one discussed above, and may become observable if $q_F \lambda_c \ll 1$, a condition which is definitely realistic: The value of the filling factor $q_F/k_F$ can indeed be adjusted by an external gate bias, and the recent developments in contact technology allow to realize extremely thin contacts, such as finger-shaped electrodes of about $10 {\rm nm}$, or superconducting STM tips. Here we show that in this case the Josephson current exhibits interesting novel features.\\

\begin{figure} 
\includegraphics[width=\columnwidth]{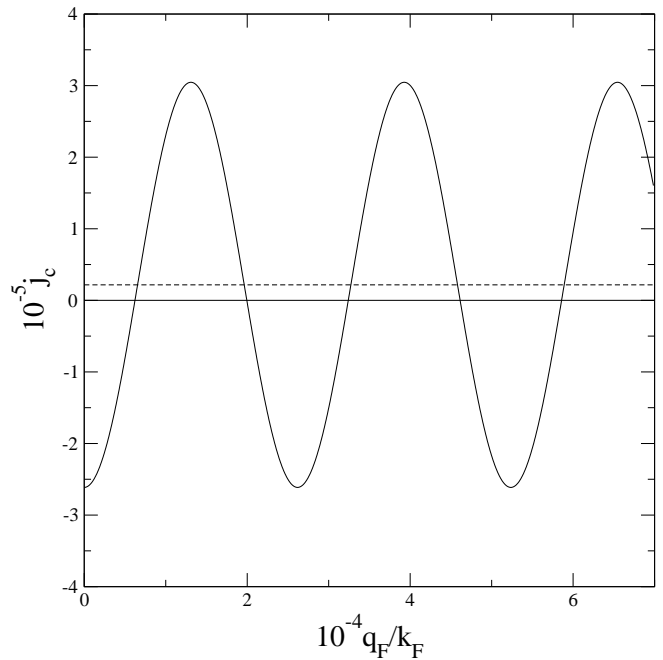}
\caption{\label{fig:4} (solid): The  dimensionless critical current $j_c$ as a function of the filling factor $q_F/k_F$ at zero temperature for a S-I-SWNT-I-S junction with length $d=6 \cdot 10^3 k_F^{-1}$. The interaction parameter is $K=0.3$. By tuning the electron filling, e.g.~with a gate voltage, the sign of the critical current can be reversed, tuning the junction from a $0$ into a $\pi$-state.
(dashed): The contribution of $A$-processes to the critical current, independent of the filling factor.}
\end{figure}

\begin{figure} 
\label{fig5}
\includegraphics[width=\columnwidth]{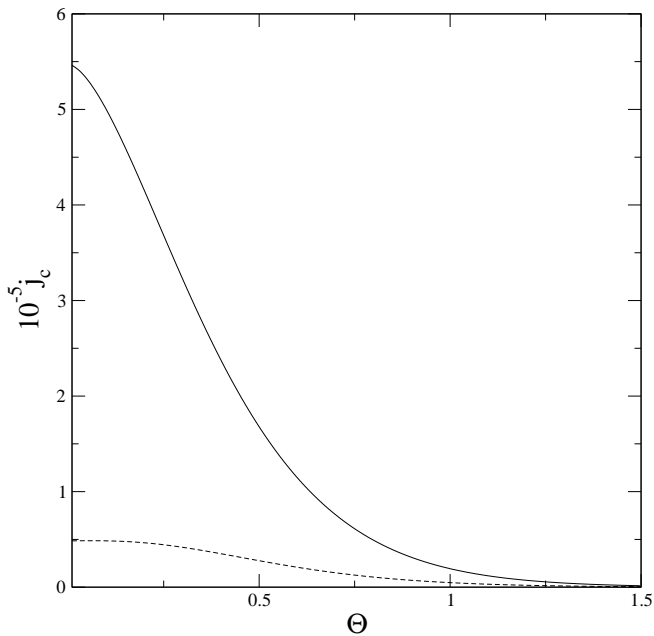}
\caption{\label{fig:5} (solid): The dimensionless critical current $j_c$ as a function of the reduced temperature $\Theta$ for a junction with length $d=6 \cdot 10^3 k_F^{-1}$, realized with a SWNT with filling factor $q_F/k_F =10^{-3}\pi/12$ and interaction parameter $K=0.3$.
(dashed): The contribution of $A$-processes to the critical current. Differently from the case of  a non-interacting wire, in a SWNT the contribution due to tunnel of pairs with antiparallel momenta ($A$ processes) is dominated by the one originating from pairs with parallel momenta ($P$ processes).}
\end{figure}

Fig.~\ref{fig:3} displays the  dimensionless Josephson critical current~$j_c$,  Eq.~(\ref{jd}), as a function of the junction length $d$ for a SWNT with interaction strength $K=0.3$   at zero temperature. We recall that the approximation of non-interacting electrons would predict a constant value for~$j_c$. In contrast, in a SWNT the strong electron interaction  leads~$j_c$ to decay with an oscillatory behavior as a function of~$d$. While the power law decay has been predicted also for usual two-channel LL, the oscillatory behavior is purely due to the four channel band structure of nanotube. Importantly, this implies that the {\it sign} of the Josephson current depends on the length of the junction, and that SWNT can be used to realize a $\pi$-state. We emphasize that this effect originates from $P$ processes;   the contribution to $j_c$ due to the A processes, described by the dashed curve in  Fig.~\ref{fig:3}, is indeed monotonous  and always positive.\\
Fig.~\ref{fig:4} shows $j_c$ for a S-I-SWNT-I-S junction with length $d= 6\cdot10^3 k_F^{-1} \sim 360\,\textrm{nm}$: the Josephson current oscillates with a period $\pi/k_F d$ as a function of the filling factor $q_F/k_F$, around a value (dashed line) which represents the contribution of $A$ processes, independent of $q_F$. By tuning $q_F$ with a gate voltage, the switching from a $0$ to a $\pi$-junction behavior can be induced.\\
Finally, Fig.~\ref{fig:5} shows the behavior of $j_c$ as a function of the dimensionless temperature $\Theta=k_B T \, d/\hbar v_F$. As expected from Eq.~(\ref{jd}), thermal fluctuations suppress the Josephson effect  at a temperature of the order of $k_B T \sim \hbar v_F/d$ (for a 100 nm long junction this corresponds to   $T=60 {}^o {\rm K}$); the figure elucidates the crucial role played by the interaction in determining the relative magnitude of $P$ processes with respect to $A$ processes: while for a non-interacting wire ($K=1$) the $P$ processes contribution vanishes, for  a SWNT ($K \simeq 0.3$) the latter dominate for a sufficiently long junction. 

\section{Conclusions}

In this paper we have investigated the dc Josephson effect in a S-I-SWNT-I-S   junction.  The effects of the electron-electron interaction on the critical current~$j_c$ have been particularly analyzed by using the Luttinger Liquid theory. We have found that~$j_c$ oscillates with a factor  $2 q_F d$, where $d$ is the junction length and $q_F$ the Fermi momentum characterizing the electron filling with respect to the band crossing point $k_F$. These oscillations are a signature of the peculiar band structure and of  the  strong electronic correlations present in SWNT. We emphasize that they would indeed not appear in non-interacting systems. Remarkably, this effect implies that ballistic SWNT can be used to realize tunable $\pi$-junctions, for the sign  of the critical current can be controlled  by varying either the filling factor or the junction length (see Figs.~\ref{fig:3} and \ref{fig:4}). The former  can be tuned through an external gate voltage. The latter can be changed for instance by moving  the superconducting tip of an STM\cite{panetal98,rodrigoetal04,kohenaetal05} along the nanotube. The typical value of $k_F$ is of the order of $20$ nm$^{-1}$, so that the predicted oscillations should be  observable  in junctions with length $d \gtrsim 100 {\rm nm}$ operating at temperature of the order of some ${}^o {\rm K}$ or below; the SWNT should have an electron filling momentum $q_F$ ranging from 0 up to a small fraction of $\lambda_c^{-1}$.

\section*{Acknowledgements}
\noindent Fruitful discussions with F. Giazotto as well as financial support from HYSWITCH EU Project are greatly acknowledged. 
\appendix

\section{$O^i_r$ correlation functions}
In this appendix we provide the expressions for the T-ordered correlation functions appearing in  the computation of the Josephson current Eq.~(\ref{jd}). The correlation functions can be written as the product of their ground state value and a thermal fluctuations contribution, which equals 1 at zero temperature. Explicitly:
\begin{equation}
\label{eq:Ocorrelationfunctions}
\begin{split}
C_r^{A}(k_F d, \xi)
=&
\big[
  \mathfrak{g}_r^{A\,(GS)}(k_F d, \xi) \, \mathfrak{g}_r^{A\,(TF)}(k_F d, \xi)
\big]^2\\
&\\
C_r^{P}(k_F d, \xi)
=&
\big|
  \mathfrak{g}_r^{P\,(GS)}(k_F d, \xi) \, \mathfrak{g}_r^{P\,(TF)}(k_F d, \xi)
\big|^2
\end{split}
\end{equation}
where:

\begin{equation} 
\begin{split}
\mathfrak{g}_r^{A\,(GS)}(k_F d, \xi)
=&
\left|
  \frac {\tilde a }{ \tilde a + z^r_K}
\right|^{\delta_A-1/2} \left(
  \frac{\tilde a}{ \tilde a+ z_1^*}
\right)^{\frac{1}{2}}
\\
&\cdot \left(
  \frac  { \tilde a+ z^r_K} { \tilde a+ {z^r_1}^*}
\right)^{\frac{1}{4}}
\end{split}
\end{equation}
\begin{equation} 
\begin{split}
\mathfrak{g}_r^{P\,(GS)}(k_Fd, \xi)
=&
\left|
  \frac {\tilde a }{ \tilde a + z^r_K}
\right|^{\delta_P}
\left(
  \frac{ \tilde a+ z^r_K}{ \tilde a+ {z^r_1}^*}
\right)^{\frac{1}{2}} \, ,
\end{split}
\end{equation}
with $ z^r_K=k_F d(i \,  r \mathrm{sign}(\xi)+|\xi|/K)$ and $\tilde a=k_Fa$. The contribution due to thermal fluctuations reads
\begin{equation}
\begin{split}
\mathfrak{g}_r^{A\,(TF)}(k_F d, \xi)
=&
\left|
  \frac{K\pi\Theta  z^r_K}{\sin(K\pi\Theta  z^r_K)}
\right|^{\delta_A -1/2}
\left(
  \frac{K\pi\Theta  z^r_K}{\sin(K\pi\Theta  z^r_K)}
\right)^{-\frac{1}{4}}\\
&\cdot
\left(
  \frac{\pi\Theta {z^r_1}^*}{\sin(\pi\Theta  {z^r_1}^*)}
\right)^{\frac{3}{4}}
\end{split}
\end{equation}
\begin{equation}
\begin{split}
\mathfrak{g}_r^{P\,(TF)}(k_F d, \xi)
=&
\left|
  \frac{K\pi\Theta  z^r_K}{\sin(K\pi\Theta  z^r_K)}
\right|^{\delta_P}
\left(
  \frac{K\pi\Theta  z^r_K}{\sin(K\pi\Theta  z^r_K)}
\right)^{-\frac{1}{2}}\\
&\cdot
\left(
  \frac{\pi\Theta  {z^r_1}^*}{\sin(\pi\Theta  {z^r_1}^*)}
\right)^{\frac{1}{2}} \, \, .
\end{split}
\end{equation}
The cut-off $\tilde{a}$ renormalizes the bare tunneling amplitude $\mathcal{T}$ in different ways for $P$ and $A$ processes. In particular, one has $\mathcal{T} \rightarrow \mathcal{T}_i =\mathcal{T} (k_F a)^{2(\delta_i -1)}$, with $i=A,P$. Typically $\tilde{a}  \lesssim 1$ (here we have chosen $\tilde{a} =0.5$).



\bibliography{db1}

\end{document}